\documentclass{vgtc}                          % final (conference style)

\ifpdf%                                % if we use pdflatex
\pdfoutput=1\relax                   % create PDFs from pdfLaTeX
\usepackage{graphicx}                % allow us to embed graphics files
\DeclareGraphicsExtensions{.pdf,.png,.jpg,.jpeg} % for pdflatex we expect .pdf, .png, or .jpg files
\else%                                 % else we use pure latex
\usepackage{graphicx}                % allow us to embed graphics files
\DeclareGraphicsExtensions{.eps}     % for pure latex we expect eps files
\fi%

\graphicspath{{figures/}{pictures/}{images/}{./}} % where to search for the images

\usepackage{microtype}                 % use micro-typography (slightly more compact, better to read)
\PassOptionsToPackage{warn}{textcomp}  % to address font issues with \textrightarrow
\usepackage{textcomp}                  % use better special symbols
\usepackage{mathptmx}                  % use matching math font
\usepackage{times}                     % we use Times as the main font
\usepackage{tabu}                      % only used for the table example
\usepackage{booktabs}                  % only used for the table example
\usepackage{enumitem,subcaption}
\usepackage[numbers,square,sort]{natbib}
\renewcommand{\quote}[1]{\textit{``#1''}}
\newcommand{\citeg}[1]{\cite[e.g.,][]{#1}}

\onlineid{0}

\vgtccategory{Research}

\vgtcinsertpkg
\usepackage[nameinlink,capitalise]{cleveref}

\title{Trust in Prediction Models: a Mixed-Methods Pilot Study\\on the Impact of Domain Expertise}

\author{Jeroen Ooge\thanks{e-mail: jeroen.ooge@kuleuven.be}%
  \and Katrien Verbert\thanks{e-mail: katrien.verbert@kuleuven.be}}
\affiliation{\scriptsize Department of Computer Science \\ KU Leuven}

\abstract{
People's trust in prediction models can be affected by many factors, including domain expertise like knowledge about the application domain and experience with predictive modelling. However, to what extent and why domain expertise impacts people's trust is not entirely clear. In addition, accurately measuring people's trust remains challenging. We share our results and experiences of an exploratory pilot study in which four people experienced with predictive modelling systematically explore a visual analytics system with an unknown prediction model. Through a mixed-methods approach involving Likert-type questions and a semi-structured interview, we investigate how people's trust evolves during their exploration, and we distil six themes that affect their trust in the prediction model. Our results underline the multi-faceted nature of trust, and suggest that domain expertise alone cannot fully predict people's trust perceptions.
} % end of abstract

\begin{document}
  \maketitle
  
  \section{Introduction}
  Intelligent systems like visual analytics systems are increasingly incorporating artificial intelligence to support end-users in decision-making \cite{lu_stateart_2017,endert_state_2017,hohman_visual_2019}. To make well-informed decisions, it is vital that people appropriately trust the underlying models \cite{gunning_darpas_2019,han_beyond_2020}. Therefore, lots of research has been dedicated to trust in human-computer interaction \citeg{dzindolet_role_2003,cramer_effects_2008,papenmeier_how_2019,uggirala_measurement_2004}, specifically in information visualisation~\cite{chatzimparmpas_state_2020} and explainable artificial intelligence \citeg{yang_how_2020,ribeiro_why_2016,bussone_role_2015}.
  
  However, trust is a slippery concept because it is related to many factors \cite{hoff_trust_2015}. One example is domain expertise, which can refer to artificial intelligence or the application domain in question. Previous studies have shown that both facets can influence people's trust in an automated system \cite{bussone_role_2015,nourani_role_2020,hoff_trust_2015}. Other factors that might affect trust include the way in which information is visualised \cite{mayr_trust_2019}, age \cite{knowles_older_2018}, uncertainty \cite{sacha_role_2016}, cognitive load \cite{bernhaupt_effects_2017}, model accuracy \cite{yin_understanding_2019}, algorithmic transparency \cite{kizilcec_how_2016}, and the point in time on which the intelligent system is used \cite{master_measurement_2005,nourani_role_2020,holliday_user_2016,mohseni_trust_2020,nourani_investigating_2020}. As a consequence of this long list of influential factors, measuring trust is very challenging. Researchers have proposed Likert-scales that capture people's trust in an automated system \citeg{gulati_design_2019,bernhaupt_modelling_2017,jian_foundations_2000,madsen_measuring_2000}, often inspired by the psychological literature on trust relations between humans \cite{hoffman_trust_2013}. However, there is still debate about these scales' validity, and even about whether explainable artificial intelligence should focus on trust in the first place \cite{davis_measure_2020}.
  
  In this paper, we share our results and experiences of a mixed-methods pilot study with four participants who are familiar with predictive modelling, and active in agrifood domains. Our research contribution is threefold:
  \begin{enumerate}[noitemsep,leftmargin=*]
    \item To measure people's expertise and trust in a prediction model, we propose a mixed-methods approach that goes beyond using single Likert-type questions, yet remains feasible in real-life studies;
    \item We illustrate that only knowing people's expertise in predictive modelling does not suffice to predict their trust in a prediction model;
    \item By thematically analysing the transcripts of semi-structured interviews, we extract six factors that might influence people's trust in a prediction model.
  \end{enumerate}
  
  \section{Research Motivation}
  Research on trust in intelligent systems often subdivides people into those who are familiar with a certain topic (``experts''), and those who are not (``non-experts'' or ``laypeople'') \citeg{nourani_role_2020,bernhaupt_effects_2017}. The research goal is then to find differences between, and similarities within those groups. We were interested in the latter, particularly in whether people experienced with predictive modelling have similar trust perceptions when they explore a visual analytics system without knowing the underlying prediction model. Inspired by studies on trust evolution over time \cite{master_measurement_2005,nourani_role_2020,holliday_user_2016,mohseni_trust_2020,nourani_investigating_2020}, we decided to show people increasingly more visual information about a prediction outcome, and to capture their trust evolution. Specifically, our research questions were as follows:
  \begin{itemize}[noitemsep,leftmargin=*]
    \item \textbf{RQ1}. Do people experienced with predictive modelling have similar trust levels and evolutions for an unknown prediction model?
    \item \textbf{RQ2}. What influences trust in an unknown prediction model for people experienced with predictive modelling?
  \end{itemize}
  To make fair comparisons, we needed participants with similar backgrounds. We chose to target people in agrifood because research on trust and uncertainty visualisation is limited in this domain \cite{gutierrez2019review}.
  
  \section{Materials and Methods}
  This section presents how we conducted our study. We first describe our visual analytics system and overall study design. Then, we provide more details on how we measured expertise and trust.
  
  \subsection{Visual Analytics System}
\begin{figure*}[tb]
    \centering
    \begin{subfigure}{.495\textwidth}
      \includegraphics[width=\linewidth]{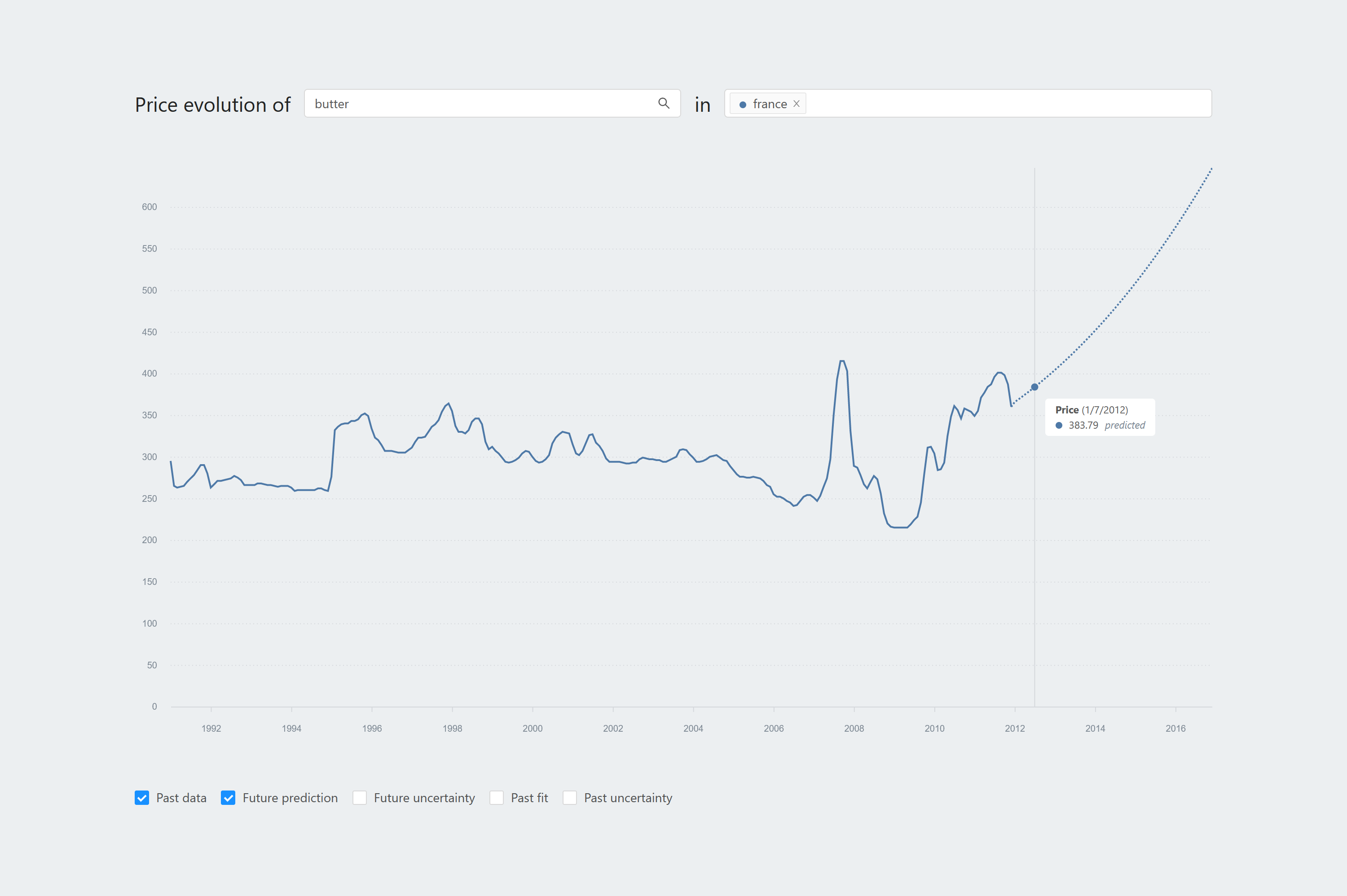}
      \caption{Scenario 1: the future prediction for one country is visualised as a dashed line.}
      \label{fig:dashboard1}
    \end{subfigure}\hfill%
    \begin{subfigure}{.495\textwidth}
      \includegraphics[width=\linewidth]{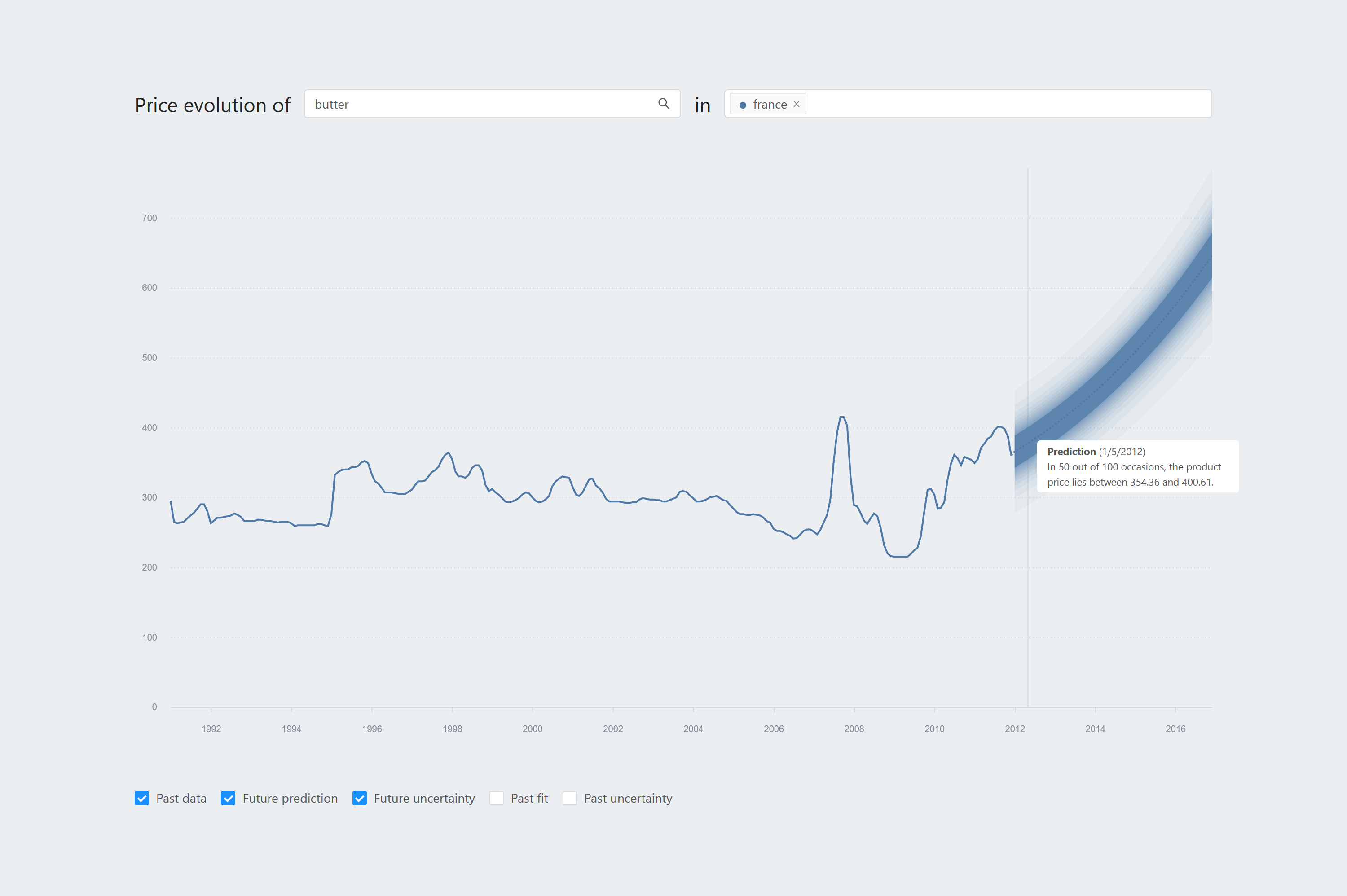}
      \caption{Scenario 2: the future uncertainty for one country is visualised as fans.}
      \label{fig:dashboard2}
    \end{subfigure}
    \vskip\baselineskip
    \begin{subfigure}{.495\textwidth}
      \includegraphics[width=\linewidth]{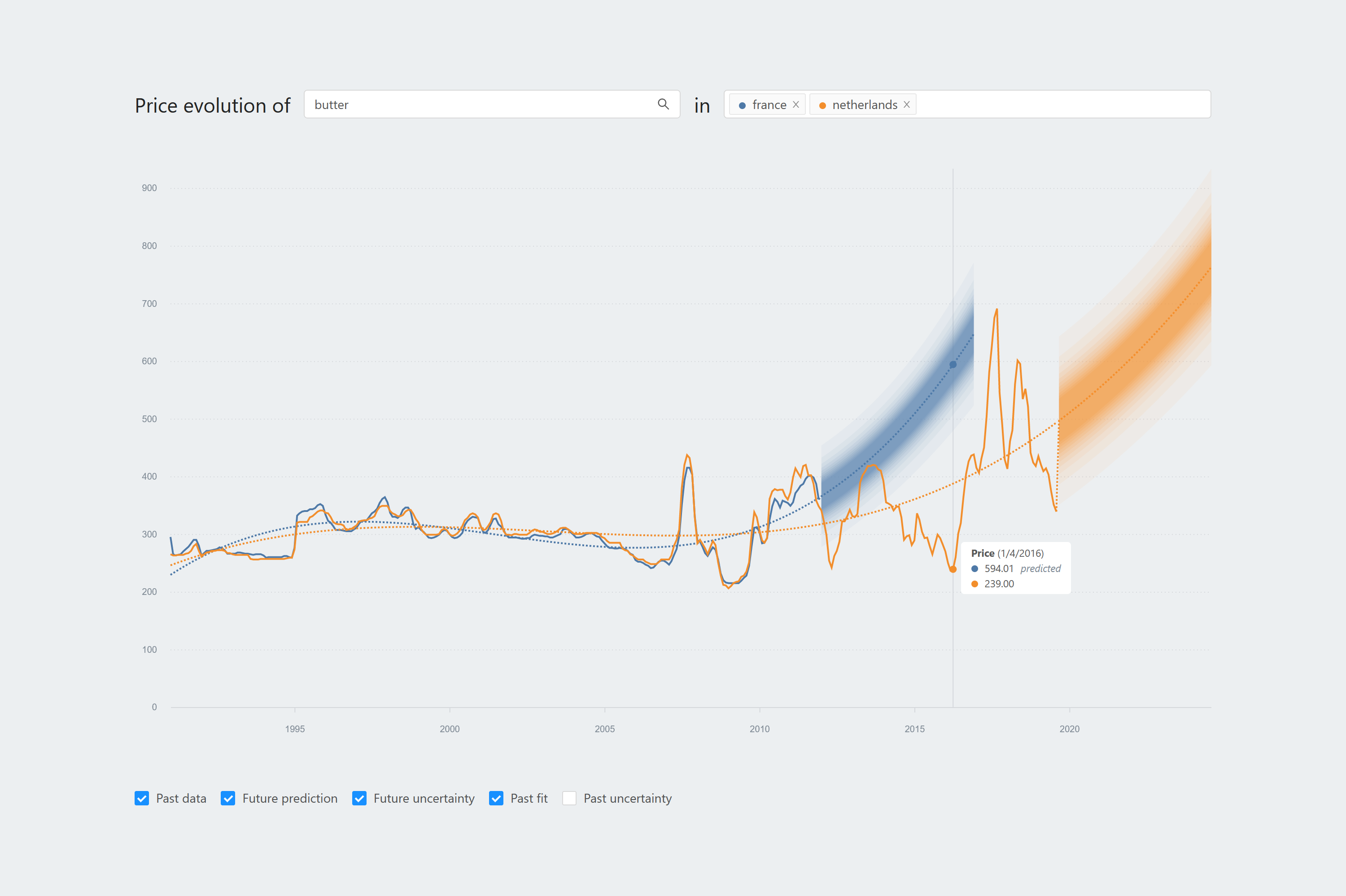}
      \caption{Scenario 7: the past fit for two countries is visualised as dashed lines.}
      \label{fig:dashboard3}
    \end{subfigure}\hfill%
    \begin{subfigure}{.495\textwidth}
      \includegraphics[width=\linewidth]{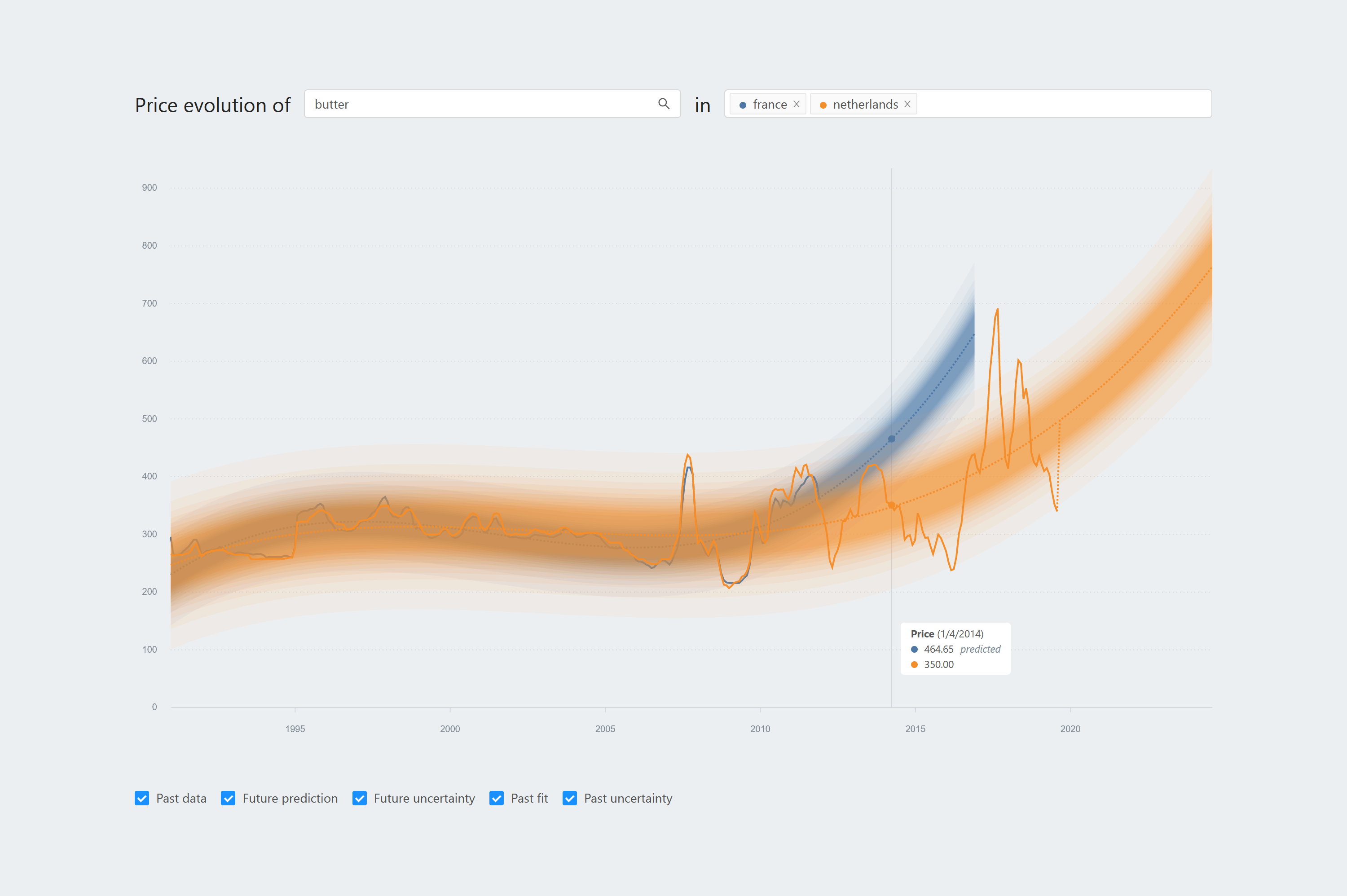}
      \caption{Scenario 8: the past uncertainty for two countries is visualised as fans.}
      \label{fig:dashboard4}
    \end{subfigure}
    \caption{Screenshots of our visual analytics system, illustrating the enabled visual components in four out of eight scenarios.}
    \label{fig:dashboard}
  \end{figure*}

  We developed a simple visual analytics system for exploring the price evolution of food products in various European countries. For each country, we fit a third-degree polynomial to the available past data with linear regression and least-squares estimation, used a five-year extrapolation as prediction, and computed prediction intervals at levels 50 to 99 with increments of five. Obviously, more sophisticated techniques for forecasting time series exist; we used linear regression only for illustration purposes. \autoref{fig:dashboard} shows our dashboard with at the bottom five checkboxes that enable visual components related to the prediction outcome and model: \textit{Past data}, \textit{Future prediction}, \textit{Future uncertainty}, \textit{Past fit}, and \textit{Past uncertainty}. The past data were visualised as a full line, future prediction and past fit as dashed lines, and the uncertainty as stacked coloured bands (also called fans). Our system was built with Meteor, React and D3, and is available at \url{https://github.com/BigDataGrapes-EU/product-prices-public}.
  
  \subsection{Study Design}
  With our visual analytics system, we conducted a case-study focused on people's trust in the prediction model, following a mixed-methods approach. In particular, we collected data on trust quantitatively with Likert-type questions, and qualitatively with a semi-structured interview. Four people participated; they were collected via email by an industrial partner in the scope of a larger research project.
  
  Our study was conducted online and was structured as follows. First, participants briefly introduced themselves and their experience with predictive modelling. Then, we demonstrated our visual analytics system, showing the past data and future prediction for a single country (Scenario~1, \autoref{fig:dashboard1}) without revealing the underlying prediction model. Next, participants went through eight scenarios by enabling the \textit{Future prediction}, \textit{Future uncertainty}, \textit{Past fit} and \textit{Past uncertainty} checkboxes in our dashboard one by one, respectively, first for a setting with one country (Scenarios~1--4), and then for a setting with two countries (Scenarios~5--8). \autoref{fig:dashboard} shows some representative screenshots. In each scenario, we first asked participants to interact with the visualisation and to describe their initial impressions of the visualisation in a think-aloud fashion (\textit{Explore the new component in the visualisation. Explain what you see. What grabs your attention?}). Next, we asked them about their trust in the prediction model through open-ended questions (\textit{Do you trust the prediction model? Which parts of the visualisation made you say that?}) and four Likert-type questions (see \autoref{sec:measuringtrust}). Finally, after completing all scenarios, participants reported their familiarity with four concepts related to predictive modelling (see \autoref{sec:measuringexpertise}). In the post-study discussion, we asked participants how they experienced the study, and stressed the illustrative nature of our prediction model.
  
  \subsection{Measuring Trust}
  \label{sec:measuringtrust}
  To quantitatively measure people's trust in the prediction model, we drew inspiration from a trust scale for automated systems by Jian et al. \cite{jian_foundations_2000}, which does not consider willingness to act upon a system's advice, contrary to Madsen and Gregor's trust definition~\cite{madsen_measuring_2000}. We considered it unfeasible for participants to answer the twelve items in the original scale eight times, so we selected and adapted four items that seemed most appropriate for trust in prediction models:
  \begin{enumerate}[noitemsep,leftmargin=*]
    \item (R) I am suspicious of the prediction model's outputs;
    \item I am confident in the prediction model;
    \item I can trust the prediction model;
    \item (R) The prediction model is deceptive.
  \end{enumerate}
  Items were rated on a 7-point range, and reverse-scored~(R) when necessary. The four scores between 0 (not at all) and 6 (extremely) were then summed, resulting in a final trust score between 0 and 24.
  
  \subsection{Measuring Expertise}
  \label{sec:measuringexpertise}
  To verify that participants were experienced with predictive regression analysis, we combined self-reported data and indirect familiarity indications. First, participants self-reported their familiarity with \textit{prediction interval}, \textit{linear regression}, \textit{time series prediction}, and \textit{fan chart} through checkboxes ``I know the word''~(K), ``I often use it''~(U) and ``I can explain it''~(E). For each concept, we assigned a score between 0 (not familiar at all) and 5 (completely familiar) based on their answers (K = 1, K+U = 3, K+E = 4, K+U+E = 5), and the average $S_1$ served as a final estimate for self-reported familiarity. Second, we scored participants' expertise between 0 and 5 based on their background ($S_2$) and use of jargon related to statistics or predictive modelling during the interview ($S_3$). Finally, we considered participants to be sufficiently experienced with predictive regression analysis if the average of their $S_1$, $S_2$ and $S_3$ scores exceeded 3.5.
  
  \section{Results}
  This section presents the results of our studies with four participants, each lasting 70--100 minutes. We first investigate the overall trust evolution over the eight scenarios to spot differences and similarities between participants. Then, we contextualise observed trends by thematically analysing \cite{braun2012thematic} the qualitative feedback collected during the semi-structured interviews.
  
  \subsection{Domain Expertise and Quantitative Trust Evolution}
  \autoref{tab:participants} shows that all participants had an expertise score of over 3.5, which confirms that they belong to our target group of people experienced with predictive modelling. Interestingly, P3 and P4 had a self-reported expertise score far below the scores based on their background and jargon use.
  
  \begin{table*}
	\centering
	\caption{Background information about the participants of our study, including their estimated expertise in predictive modelling.}
	\label{tab:participants}
	\begin{tabular}{@{}lllll@{}}
		\toprule
		\textbf{ID} & \textbf{Profession}                 & \textbf{Age} & \textbf{Country} & \textbf{Expertise ($S_1$, $S_2$, $S_3$)} \\
		\midrule
		P1           & Quality manager, analyst (industry) & 45--54       & Greece           & 4.58 (3.75, 5, 5)                      \\
		P2           & Agrifood engineer (academia)        & 45--54       & Italy            & 4.17 (4, 4, 4.5)                       \\
		P3           & Agricultural economist (academia)   & 35--44       & Italy            & 3.67 (2, 4.5, 4.5)                     \\
		P4           & Agricultural researcher (industry, academia)     & 35--44       & Greece           & 4.25 (2.75, 5, 5)\\
		\bottomrule
	\end{tabular}
	\\[5pt]\scriptsize
	All participants identified as male. Expertise scores: $S_1$ = self-reported, $S_2$ = background, $S_3$ = jargon use.
\end{table*}
  
\begin{figure}
    \centering
    \includegraphics[width=\columnwidth]{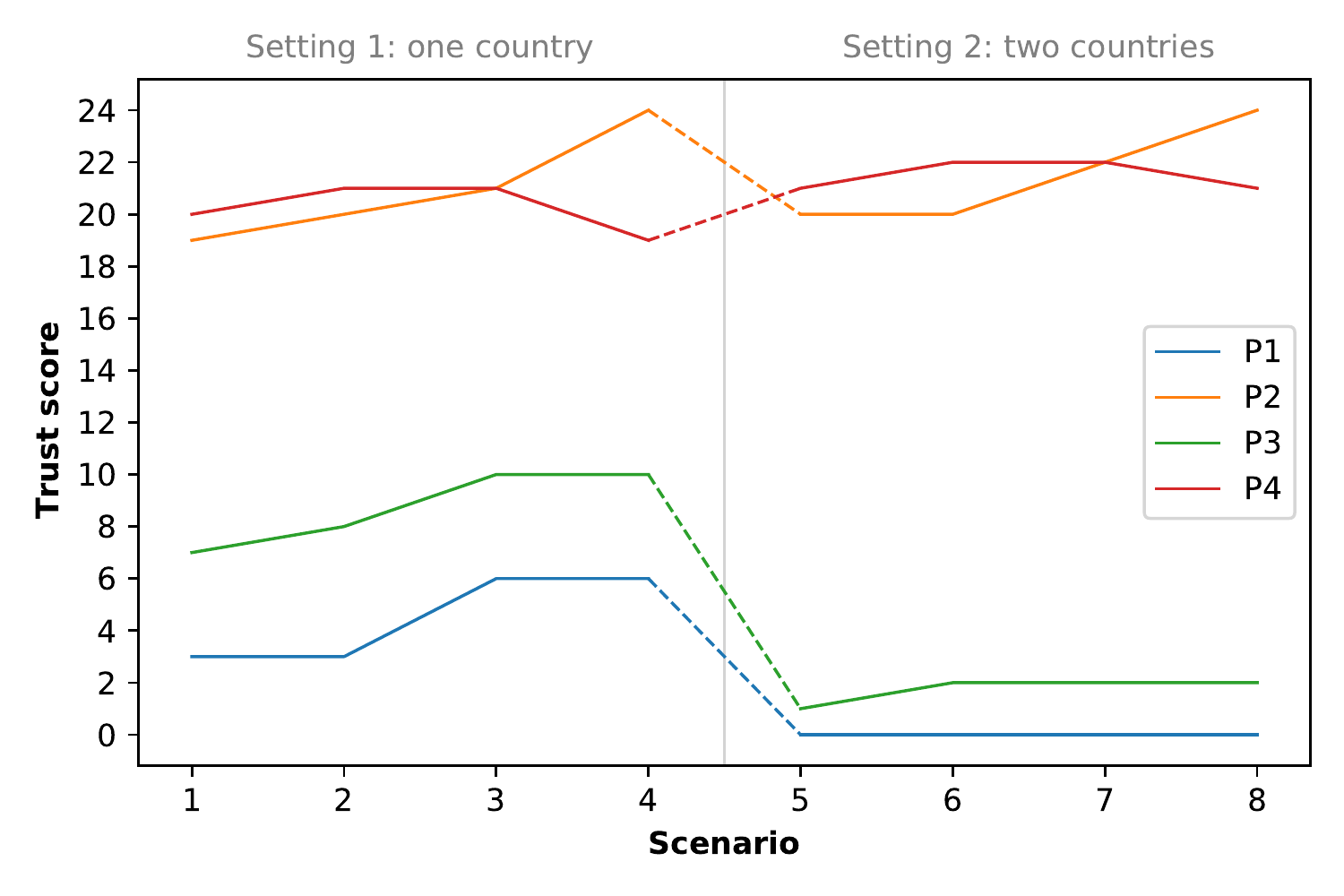}
    \caption{Participants' trust scores in the prediction model, evolving over eight scenarios. The first four scenarios show a single country's data, and the next four scenarios show two countries. Overall, P1 and P3 reported low trust scores; P2 and P4 high trust scores.}
    \label{fig:trustevolution}
  \end{figure}

  \autoref{fig:trustevolution} shows the evolution of participant's reported trust scores over all scenarios. Recall that Scenarios 1--4 showed data for one country (Setting~1), whereas Scenarios 5--8 showed data for two countries (Setting~2). Looking at the overall trust scores, there is a clear distinction between two pairs of people: P2 and P4 were trusting the prediction model, P1 and P3 were not. Thus, although all participants had comparable expertise in predictive modelling, their trust in the prediction model did not evolve similarly over the eight scenarios. Comparing Setting~1 and Setting~2, we observe two things. First, both settings show a non-decreasing trend in trust. Only P4 slightly violates this trend in the transitions of Scenarios 3 to 4, and 7 to 8; the transitions where the past uncertainty becomes visible. Second, there is a drop in trust when participants switched from Setting~1 to Setting~2. Only P4 is again the exception.
  
  \subsection{Thematic Analysis of Qualitative Trust Feedback}
  Our quantitative results alone cannot explain the trends they hold. To contextualise the observed trust evolutions, we consult our rich qualitative data, and present six trust themes that emerged during the participants' discussions, ordered by prominence.
  
  \textbf{Expectations about model outcomes (T1)}. This theme consists of two complementary parts.
  
  (1)~\textit{Expectation violation}: when participants encountered an unexpected result, their trust in the prediction model typically decreased. For example, P3 reported low trust in Scenario~1, stating: \quote{If I look at these time series, I wouldn't expect such a high slope of the curve, of the data. Theoretically I was expecting a less high slope.} Another example concerns Scenarios~5--8 (see \cref{fig:dashboard3,fig:dashboard4}), which showed two closely linked price evolutions, whereby one country's prediction diverged from the other country's corresponding past data. P1 and P3 experienced this as a strong expectation violation, leading to a huge drop in trust. Finally, in Scenarios~4 and 8, P4 expected less uncertainty in the centre of the time series and more uncertainty in the tails, leading to a slight decrease in trust.
  
  (2)~\textit{Expectation agreement}: when participants observed something that matched their expectation, their trust in the prediction model often increased. For example, P4 reported high trust in Scenario~1, stating: \quote{usually we have increas[ing] prices, not decreases [laughs], so that's why I'm more in the part that I'm trusting the prediction.} Later on, P4 repeatedly indicated that he strongly suspected the model to be a regression because the visualised data seemed to confirm that: \quote{I don't see anything strange. [...] I believe it’s just a regression model.} In the post-study discussion, P1 also hinted at this trust theme, emphasising the importance of \quote{reasonable reasoning behind what we see in the model.}
  
  \textbf{Understanding the prediction model (T2)}. P1, P2 and P3 often repeated the need to have a detailed understanding of the prediction model underlying our visual analytics system. P1 even mentioned this right from the start: \quote{In order to trust a prediction model, I need to know how it was developed.} Also in Scenario~1, P3 answered the question on whether he trusted the model with: \quote{to be honest not, because [...] I have no idea how you provide this prediction, how you calculate it and the model behind.} More on a technical level, participants expected the prediction model to consider multiple variables like consumer behaviour, and the political, climatological and economical context.
  
  \textbf{Predictions need uncertainty (T3)}. All participants received uncertainty in the prediction outcomes positively. On the one hand, P2 stated to trust the model more because the uncertainty shows a \quote{statistical consideration} and suggests that previous studies underpin the presented values. P4 agreed: \quote{the more descriptive the model becomes, and the more alternatives that it gives you, it makes you trust more. When you have just a line, you more or less, you cannot believe that things in real life are so accurate, right? [chuckles].} On the other hand, P1 and P3 considered reasonable uncertainty in the prediction a natural requirement, and stressed that uncertainty alone was not sufficient to increase their trust dramatically: \quote{it's a model that takes some reasonable uncertainty, but still I cannot trust it because I don't know how it was developed} (P1 in Scenario~2).
  
  \textbf{Developers of the prediction model (T4)}. Two participants based their trust in the prediction model not only on its outcomes, but also on its developers. In Scenario~1, P2 referred to trusting the model developers' competence: \quote{when I approach information that come[s] from an organisation or something like you, I suppose, my behaviour is to accept this evolution because I suppose that you have the competence to develop a model.} P3 argued that a model stemming from an official institution might be more reliable: \quote{If such a prediction comes from an official body like FAO or World Bank or so on, [it] could be more reliable. If [it] come[s] from a university [...], it's not an official body and it's more difficult to understand. [...] when a World Bank provides [a] prediction, it's the fruit of the convergent opinion of different practitioners and scientists.}
  
  \textbf{Data provenance (T5)}. P1, P2 and P3 indicated the importance of guaranteed accurate data and transparency about their origin. Characteristic quotes are: \quote{In order to trust a prediction model, I need to know [...] what is the raw data [in]put} (P1 in Scenario~1) and \quote{[It] could be important [...] to know in a precise way the provenance of the past data. OK, the past data derived from which database or when [did] you take this data for the past?} (P3 in the post-study discussion).
  
  \textbf{Past performance of the prediction model (T6)}. Transitioning from Scenario~2 to Scenario~3, P1 and P3 reported an increase in trust and their qualitative feedback suggests that the reason for this is the prediction model's good performance in the past: \quote{this particular model has explained reasonably, reasonably, the variation of butter price[s] throughout the decades} (P1); \quote{The model seems to fit very [well] with the data. It gives more robustness to the model} (P3).
  
  \section{Discussion and Future Work}
  This section discusses our results: not all ``experts'' have similar trust levels and evolutions (RQ1). Instead, trust in a prediction model is influenced by several interconnected themes, and people's dominant themes can evolve under new information (RQ2).
  
  \subsection{An ``Expert'' Label Does Not Say It All}
  While participants in our study all had a background in agrifood and had similar expertise with predictive modelling, their trust levels and evolutions over the eight scenarios in our study clearly differed: two participants had high trust levels, while two others distrusted the prediction model. This suggests that ``experts'' do not always respond in a completely similar way to an unknown prediction model, and that their trust levels and evolutions are subject to different factors (RQ1). Future research could investigate whether the distinction that we identified holds in a larger sample of people experienced with predictive modelling, and whether similar patterns also occur for people lacking this expertise.
  
  \subsection{Trust Is Multi-Faceted}
  The thematic analysis of our semi-structured interviews subscribes the multi-faceted nature of people's trust in a prediction model. We identified six themes (RQ2), which show that people's trust can be influenced by their expectations about the model outcomes (T1), their understanding of the prediction model (T2), the uncertainty in model outcomes (T3), the model's developers (T4), the data provenance (T5), and the model's past performance (T6).
  
  Overall, the most prominent theme was T1, which turned out to be closely connected to domain knowledge. For example, participants often noticed that the prediction model could not predict a peak in 2007, but added that this outlier could have been due to the financial crisis at that time. This suggests a strong link between domain knowledge and trust themes like T1 and T6. Thus, the question rises how prominent these themes are for people with low domain expertise, and whether misinterpretations of predictions cause mistrust or misdistrust \cite{han_beyond_2020}.
  
  Another frequently raised trust theme was T2. Intuitively, this matches with the expertise of our participants: they are well-aware of how prediction models can be built, and of their potential to mislead if applied incorrectly. P1, P2 and P3 even brought these points up literally. Two quotes clearly suggest that T2's prominence is related to the participants' background: \quote{I'm an engineer, [...] understand that I always want to see the reason behind a statement or value} (P1); and \quote{I suppose this is my research mind} (P2). This begs the question whether people less experienced with predictive modelling also require a detailed understanding of prediction models to gain trust in them.
  
  \subsection{Trust Themes Are Interconnected}
  Our six trust themes are not isolated. First, three themes are related to expectation: T1 focuses on model \textit{outcomes}, whereas T2 and T3 show that people experienced with predictive modelling expect the \textit{model} to consider multiple variables and uncertainty. The latter suggests that models lacking those qualities might invoke distrust because of expectation violation on a \textit{model} level. Second, T2--T5 are all related to transparency. This suggests that trust in a prediction model can be increased indirectly by improving model transparency, which is in line with common beliefs in the explainable artificial intelligence community \cite{dasgupta_familiarity_2017,davis_measure_2020}.
  
  \subsection{Dominant Trust Themes Can Evolve}
  The way in which participants analysed the data was key to which trust themes they covered and how their trust in the prediction model evolved over the scenarios. For P1 and P3, the initial dominant theme was T2, but later on, when the diverging lines in Scenarios~5--8 violated their expectations, T1 became most influential and their trust dropped. This negative impact of expectation violation is in line with \cite{kizilcec_how_2016}.  In contrast, P2 and P4 also noticed the diverging lines in Scenarios~5--8, but their trust response was different. P2 suggested that the differentiation might have been due to particular events in one of the countries, and added: \quote{in the previous analysis, [...] I [was] a little bit, uh [...] I supposed that you are an expert in this model, and so I [had] to believe you [...] with a suspicious idea. But in this case, after [the] previous analysis, I'm more convinced that the model derived is correct.} This indicates that P2's initial dominant trust theme was T4, and that his experience with the prediction model in Scenarios 1--4 led to quicker acceptance of the prediction outcomes in subsequent situations; the latter possibly related to expectation agreement (T1). P4 argued that the diverging lines in Scenarios~5--8 could not be directly compared, and remained confident in the prediction outputs, which he correctly identified as the extrapolation of a regression. This suggests that P4 trusted regression as an analysis technique on itself, rather than basing his trust on the particular outcomes; an observation that could add another dimension to T2 or raise a new trust theme in a larger study.
  
  \subsection{Transferability and Future Work}
  To conclude our discussion, we point out that our sample of four participants is probably too small to achieve full data saturation. Thus, larger studies with the same target group could investigate whether more trust themes emerge. It would also be interesting to investigate the transferability of our preliminary results to other domains and target groups. In particular, future work could include participants from domains like finance and healthcare, in which predictive modelling plays an important role too. Furthermore, researchers could involve people who are less experienced with predictive modelling to see whether trust themes similar to ours emerge, and if so, how their dominance correlates with people's background.
  
  \section{Lessons Learned About Methodology}
  Having presented the results of our pilot study, we also share four points of reflection concerning our study design, which might be relevant for future research focused on trust.
  
  (1)~There is a trade-off between nuance and feasibility. On the one hand, a single Likert-type question is quick and easy yet inadequate to capture the complexity of trust, especially because people might differentiate between points differently, and might interpret ``trust'' differently. On the other hand, full-fledged Likert-scales with many items can capture more nuance in people's trust perception, but are infeasible in studies that measure trust frequently as they exhaust participants. Our compromise of a Likert-scale with four items, complemented by qualitative feedback, seems reasonable to us: it provided rich data, and did not seem to entail respondent fatigue.
  
  (2)~Frequently asking the same questions can incite people to take more pronounced positions. This became especially clear in the interview with P3, who exclaimed in Scenario~3: \quote{Oh, I suppose you are working to have this question more decisive, more stronger answer. Yes! [...] I'm always more and more confident that the model is right, OK?} It is important for researchers to be vigilant of this phenomenon. Our advice for quantitative studies would be to foresee extra precautions in the study design to explain increasing or decreasing trust scores, whereas qualitative researchers could ask for extra clarification when participants start making bolder statements, or carefully intervene when participants make wrong assumptions about the study's intention. This seems particularly relevant when studying trust.
  
  (3)~The third item in our adapted trust scale (``I can trust the prediction model'', see \autoref{sec:measuringtrust}) is often the only item in other trust research. We suggest that researchers who choose to work with a condensed trust scale preserve this item too such that results can be compared with previous research.
  
  (4)~To estimate people's expertise, be it predictive modelling or something else, we argue that a mixed-methods approach is safer than fully relying on self-reported data. A first reason is the Dunning-Kruger effect \cite{dunning_chapter_2011}: laypeople might overestimate their knowledge about a certain topic, whereas experts might underestimate themselves. In our study, P3 and P4 are good examples: if we had only relied on their self-reported expertise, we would have excluded them wrongfully. A second reason is that assessing expertise based on a single question or several preset concepts (like $S_1$ in \autoref{sec:measuringexpertise}) might be too broad or too narrow, respectively. These arguments especially seem to hold for extensive topics like predictive modelling.
  
  \section{Conclusion}
  Our results subscribe the complexity of people's trust in a prediction model. Only knowing people's expertise in predictive modelling seems insufficient to predict their trust perceptions when they analyse the outcomes of an unknown prediction model. This could be due to many trust themes underlying people's perceptions, including the six trust themes that we identified. In sum, our pilot study illustrates the importance of applying a mixed-methods approach in trust studies.
  %Further research on trust remains important because P3 brought forward that a lack of trust might imply ignoring the prediction outcomes: \quote{I can use prediction data only if my trust on this data [information about the prediction model] is full. [...] Otherwise I think it's not useful at all: I couldn't use it [...] for a practical or a professional use, like the support for the country or the region, for specific policy, and so on.} 
  
  % Details on which components fostered (dis)trust or understanding and why that was the case will be provided in future work.
  
  \acknowledgments{This work was supported by the Research Foundation-Flanders (FWO, grant G0A3319N), the Slovenian Research Agency (grant ARRS-N2-0101), and the \emph{European Union's Horizon 2020} research and innovation program: BigDataGrapes Project (grant 780751). We thank all four participants for their insightful feedback.}

  \bibliographystyle{abbrv-doi}

  \bibliography{references}

\begin{thebibliography}{10}

\bibitem{braun2012thematic}
V.~Braun and V.~Clarke.
\newblock Thematic analysis.
\newblock In H.~Cooper, P.~M. Camic, D.~L. Long, A.~T. Panter, D.~Rindskopf,
  and K.~J. Sher, eds., {\em APA handbook of research methods in psychology,
  Vol. 2: Research designs}, pp. 57--71. American Psychological Association,
  2012.

\bibitem{bussone_role_2015}
A.~Bussone, S.~Stumpf, and D.~O'Sullivan.
\newblock The {Role} of {Explanations} on {Trust} and {Reliance} in {Clinical}
  {Decision} {Support} {Systems}.
\newblock In {\em 2015 {International} {Conference} on {Healthcare}
  {Informatics}}, pp. 160--169. IEEE, Dallas, TX, USA, Oct. 2015. doi: {{%
10\hspace{.1pt}\discretionary{.}{%
}{.}\hspace{.4pt}1109\discretionary{/}{%
}{/}ICHI\hspace{.1pt}\discretionary{.}{%
}{.}\hspace{.4pt}2015\hspace{.1pt}\discretionary{.}{%
}{.}\hspace{.4pt}26}}


\bibitem{chatzimparmpas_state_2020}
A.~Chatzimparmpas, R.~Martins, I.~Jusufi, K.~Kucher, F.~Rossi, and A.~Kerren.
\newblock The {State} of the {Art} in {Enhancing} {Trust} in {Machine}
  {Learning} {Models} with the {Use} of {Visualizations}.
\newblock {\em Computer Graphics Forum}, 39(3):713--756, 2020. doi: {{%
10\hspace{.1pt}\discretionary{.}{%
}{.}\hspace{.4pt}1111\discretionary{/}{%
}{/}cgf\hspace{.1pt}\discretionary{.}{%
}{.}\hspace{.4pt}14034}}


\bibitem{cramer_effects_2008}
H.~Cramer, V.~Evers, S.~Ramlal, M.~van Someren, L.~Rutledge, N.~Stash,
  L.~Aroyo, and B.~Wielinga.
\newblock The effects of transparency on trust in and acceptance of a
  content-based art recommender.
\newblock {\em User Modeling and User-Adapted Interaction}, 18(5):455--496,
  Nov. 2008. doi: {{%
10\hspace{.1pt}\discretionary{.}{%
}{.}\hspace{.4pt}1007\discretionary{/}{%
}{/}s11257\discretionary{%
}{-}{-}008\discretionary{%
}{-}{-}9051\discretionary{%
}{-}{-}3}}


\bibitem{dasgupta_familiarity_2017}
A.~Dasgupta, J.-Y. Lee, R.~Wilson, R.~A. Lafrance, N.~Cramer, K.~Cook, and
  S.~Payne.
\newblock Familiarity {Vs} {Trust}: {A} {Comparative} {Study} of {Domain}
  {Scientists}' {Trust} in {Visual} {Analytics} and {Conventional} {Analysis}
  {Methods}.
\newblock {\em IEEE Transactions on Visualization and Computer Graphics},
  23(1):271--280, Jan. 2017. doi: {{%
10\hspace{.1pt}\discretionary{.}{%
}{.}\hspace{.4pt}1109\discretionary{/}{%
}{/}TVCG\hspace{.1pt}\discretionary{.}{%
}{.}\hspace{.4pt}2016\hspace{.1pt}\discretionary{.}{%
}{.}\hspace{.4pt}2598544}}


\bibitem{davis_measure_2020}
B.~Davis, M.~Glenski, W.~Sealy, and D.~Arendt.
\newblock Measure {Utility}, {Gain} {Trust}: {Practical} {Advice} for {XAI}
  {Researchers}.
\newblock In {\em 2020 {IEEE} {Workshop} on {TRust} and {EXpertise} in {Visual}
  {Analytics} ({TREX})}, pp. 1--8, Oct. 2020. doi: {{%
10\hspace{.1pt}\discretionary{.}{%
}{.}\hspace{.4pt}1109\discretionary{/}{%
}{/}TREX51495\hspace{.1pt}\discretionary{.}{%
}{.}\hspace{.4pt}2020\hspace{.1pt}\discretionary{.}{%
}{.}\hspace{.4pt}00005}}


\bibitem{dunning_chapter_2011}
D.~Dunning.
\newblock Chapter five - {The} {Dunning}–{Kruger} {Effect}: {On} {Being}
  {Ignorant} of {One}'s {Own} {Ignorance}.
\newblock In J.~M. Olson and M.~P. Zanna, eds., {\em Advances in {Experimental}
  {Social} {Psychology}}, vol.~44, pp. 247--296. Academic Press, Jan. 2011.
  doi: {{%
10\hspace{.1pt}\discretionary{.}{%
}{.}\hspace{.4pt}1016\discretionary{/}{%
}{/}B978\discretionary{%
}{-}{-}0\discretionary{%
}{-}{-}12\discretionary{%
}{-}{-}385522\discretionary{%
}{-}{-}0\hspace{.1pt}\discretionary{.}{%
}{.}\hspace{.4pt}00005\discretionary{%
}{-}{-}6}}


\bibitem{dzindolet_role_2003}
M.~T. Dzindolet, S.~A. Peterson, R.~A. Pomranky, L.~G. Pierce, and H.~P. Beck.
\newblock The role of trust in automation reliance.
\newblock {\em International Journal of Human-Computer Studies},
  58(6):697--718, June 2003. doi: {{%
10\hspace{.1pt}\discretionary{.}{%
}{.}\hspace{.4pt}1016\discretionary{/}{%
}{/}S1071\discretionary{%
}{-}{-}5819\discretionary{%
}{(}{(}03\discretionary{)}{%
}{)}00038\discretionary{%
}{-}{-}7}}


\bibitem{endert_state_2017}
A.~Endert, W.~Ribarsky, C.~Turkay, B.~W. Wong, I.~Nabney, I.~D. Blanco, and
  F.~Rossi.
\newblock The {State} of the {Art} in {Integrating} {Machine} {Learning} into
  {Visual} {Analytics}: {Integrating} {Machine} {Learning} into {Visual}
  {Analytics}.
\newblock {\em Computer Graphics Forum}, 36(8):458--486, Dec. 2017. doi: {{%
10\hspace{.1pt}\discretionary{.}{%
}{.}\hspace{.4pt}1111\discretionary{/}{%
}{/}cgf\hspace{.1pt}\discretionary{.}{%
}{.}\hspace{.4pt}13092}}


\bibitem{bernhaupt_modelling_2017}
S.~Gulati, S.~Sousa, and D.~Lamas.
\newblock Modelling {Trust}: {An} {Empirical} {Assessment}.
\newblock In R.~Bernhaupt, G.~Dalvi, A.~Joshi, D.~K.~Balkrishan, J.~O’Neill,
  and M.~Winckler, eds., {\em Human-{Computer} {Interaction} – {INTERACT}
  2017}, vol. 10516, pp. 40--61. Springer International Publishing, Cham, 2017.
\newblock Series Title: Lecture Notes in Computer Science. doi: {{%
10\hspace{.1pt}\discretionary{.}{%
}{.}\hspace{.4pt}1007\discretionary{/}{%
}{/}978\discretionary{%
}{-}{-}3\discretionary{%
}{-}{-}319\discretionary{%
}{-}{-}68059\discretionary{%
}{-}{-}0\_3}}


\bibitem{gulati_design_2019}
S.~Gulati, S.~Sousa, and D.~Lamas.
\newblock Design, development and evaluation of a human-computer trust scale.
\newblock {\em Behaviour \& Information Technology}, 38(10):1004--1015, Oct.
  2019. doi: {{%
10\hspace{.1pt}\discretionary{.}{%
}{.}\hspace{.4pt}1080\discretionary{/}{%
}{/}0144929X\hspace{.1pt}\discretionary{.}{%
}{.}\hspace{.4pt}2019\hspace{.1pt}\discretionary{.}{%
}{.}\hspace{.4pt}1656779}}


\bibitem{gunning_darpas_2019}
D.~Gunning and D.~Aha.
\newblock {DARPA}’s {Explainable} {Artificial} {Intelligence} ({XAI})
  {Program}.
\newblock {\em AI Magazine}, 40(2):44--58, June 2019.
\newblock Number: 2. doi: {{%
10\hspace{.1pt}\discretionary{.}{%
}{.}\hspace{.4pt}1609\discretionary{/}{%
}{/}aimag\hspace{.1pt}\discretionary{.}{%
}{.}\hspace{.4pt}v40i2\hspace{.1pt}\discretionary{.}{%
}{.}\hspace{.4pt}2850}}


\bibitem{gutierrez2019review}
F.~Guti{\'e}rrez, N.~N. Htun, F.~Schlenz, A.~Kasimati, and K.~Verbert.
\newblock A review of visualisations in agricultural decision support systems:
  An hci perspective.
\newblock {\em Computers and Electronics in Agriculture}, 163:104844, 2019.

\bibitem{han_beyond_2020}
W.~Han and H.-J. Schulz.
\newblock Beyond {Trust} {Building} — {Calibrating} {Trust} in {Visual}
  {Analytics}.
\newblock In {\em 2020 {IEEE} {Workshop} on {TRust} and {EXpertise} in {Visual}
  {Analytics} ({TREX})}, pp. 9--15. IEEE, Salt Lake City, UT, USA, Oct. 2020.
  doi: {{%
10\hspace{.1pt}\discretionary{.}{%
}{.}\hspace{.4pt}1109\discretionary{/}{%
}{/}TREX51495\hspace{.1pt}\discretionary{.}{%
}{.}\hspace{.4pt}2020\hspace{.1pt}\discretionary{.}{%
}{.}\hspace{.4pt}00006}}


\bibitem{hoff_trust_2015}
K.~A. Hoff and M.~Bashir.
\newblock Trust in {Automation}: {Integrating} {Empirical} {Evidence} on
  {Factors} {That} {Influence} {Trust}.
\newblock {\em Human Factors: The Journal of the Human Factors and Ergonomics
  Society}, 57(3):407--434, May 2015. doi: {{%
10\hspace{.1pt}\discretionary{.}{%
}{.}\hspace{.4pt}1177\discretionary{/}{%
}{/}0018720814547570}}


\bibitem{hoffman_trust_2013}
R.~R. Hoffman, M.~Johnson, J.~M. Bradshaw, and A.~Underbrink.
\newblock Trust in {Automation}.
\newblock {\em IEEE Intelligent Systems}, 28(1):84--88, Jan. 2013. doi: {{%
10\hspace{.1pt}\discretionary{.}{%
}{.}\hspace{.4pt}1109\discretionary{/}{%
}{/}MIS\hspace{.1pt}\discretionary{.}{%
}{.}\hspace{.4pt}2013\hspace{.1pt}\discretionary{.}{%
}{.}\hspace{.4pt}24}}


\bibitem{hohman_visual_2019}
F.~Hohman, M.~Kahng, R.~Pienta, and D.~H. Chau.
\newblock Visual {Analytics} in {Deep} {Learning}: {An} {Interrogative}
  {Survey} for the {Next} {Frontiers}.
\newblock {\em IEEE Transactions on Visualization and Computer Graphics},
  25(8):2674--2693, Aug. 2019. doi: {{%
10\hspace{.1pt}\discretionary{.}{%
}{.}\hspace{.4pt}1109\discretionary{/}{%
}{/}TVCG\hspace{.1pt}\discretionary{.}{%
}{.}\hspace{.4pt}2018\hspace{.1pt}\discretionary{.}{%
}{.}\hspace{.4pt}2843369}}


\bibitem{holliday_user_2016}
D.~Holliday, S.~Wilson, and S.~Stumpf.
\newblock User {Trust} in {Intelligent} {Systems}: {A} {Journey} {Over} {Time}.
\newblock In {\em Proceedings of the 21st {International} {Conference} on
  {Intelligent} {User} {Interfaces}}, pp. 164--168. ACM, Sonoma California USA,
  Mar. 2016. doi: {{%
10\hspace{.1pt}\discretionary{.}{%
}{.}\hspace{.4pt}1145\discretionary{/}{%
}{/}2856767\hspace{.1pt}\discretionary{.}{%
}{.}\hspace{.4pt}2856811}}


\bibitem{jian_foundations_2000}
J.-Y. Jian, A.~M. Bisantz, and C.~G. Drury.
\newblock Foundations for an {Empirically} {Determined} {Scale} of {Trust} in
  {Automated} {Systems}.
\newblock {\em International Journal of Cognitive Ergonomics}, 4(1):53--71,
  Mar. 2000.
\newblock Publisher: Routledge \_eprint:
  https://doi.org/10.1207/S15327566IJCE0401\_04. doi: {{%
10\hspace{.1pt}\discretionary{.}{%
}{.}\hspace{.4pt}1207\discretionary{/}{%
}{/}S15327566IJCE0401\_04}}


\bibitem{kizilcec_how_2016}
R.~F. Kizilcec.
\newblock How {Much} {Information}?: {Effects} of {Transparency} on {Trust} in
  an {Algorithmic} {Interface}.
\newblock In {\em Proceedings of the 2016 {CHI} {Conference} on {Human}
  {Factors} in {Computing} {Systems}}, pp. 2390--2395. ACM, San Jose California
  USA, May 2016. doi: {{%
10\hspace{.1pt}\discretionary{.}{%
}{.}\hspace{.4pt}1145\discretionary{/}{%
}{/}2858036\hspace{.1pt}\discretionary{.}{%
}{.}\hspace{.4pt}2858402}}


\bibitem{knowles_older_2018}
B.~Knowles and V.~L. Hanson.
\newblock Older {Adults}’ {Deployment} of ‘{Distrust}’.
\newblock {\em ACM Transactions on Computer-Human Interaction}, 25(4):1--25,
  Sept. 2018. doi: {{%
10\hspace{.1pt}\discretionary{.}{%
}{.}\hspace{.4pt}1145\discretionary{/}{%
}{/}3196490}}


\bibitem{lu_stateart_2017}
Y.~Lu, R.~Garcia, B.~Hansen, M.~Gleicher, and R.~Maciejewski.
\newblock The {State}‐of‐the‐{Art} in {Predictive} {Visual} {Analytics}.
\newblock {\em Computer Graphics Forum}, 36(3):539--562, June 2017. doi: {{%
10\hspace{.1pt}\discretionary{.}{%
}{.}\hspace{.4pt}1111\discretionary{/}{%
}{/}cgf\hspace{.1pt}\discretionary{.}{%
}{.}\hspace{.4pt}13210}}


\bibitem{madsen_measuring_2000}
M.~Madsen and S.~Gregor.
\newblock Measuring human-computer trust.
\newblock In {\em 11th australasian conference on information systems},
  vol.~53, pp. 6--8, 2000.

\bibitem{master_measurement_2005}
R.~Master, X.~Jiang, M.~T. Khasawneh, S.~R. Bowling, L.~Grimes, A.~K.
  Gramopadhye, and B.~J. Melloy.
\newblock Measurement of trust over time in hybrid inspection systems.
\newblock {\em Human Factors and Ergonomics in Manufacturing}, 15(2):177--196,
  2005. doi: {{%
10\hspace{.1pt}\discretionary{.}{%
}{.}\hspace{.4pt}1002\discretionary{/}{%
}{/}hfm\hspace{.1pt}\discretionary{.}{%
}{.}\hspace{.4pt}20021}}


\bibitem{mayr_trust_2019}
E.~Mayr, N.~Hynek, S.~Salisu, and F.~Windhager.
\newblock Trust in {Information} {Visualization}.
\newblock {\em EuroVis Workshop on Trustworthy Visualization (TrustVis)}, p. 5
  pages, 2019.
\newblock Artwork Size: 5 pages ISBN: 9783038680918 Publisher: The Eurographics
  Association Version Number: 025-029. doi: {{%
10\hspace{.1pt}\discretionary{.}{%
}{.}\hspace{.4pt}2312\discretionary{/}{%
}{/}TRVIS\hspace{.1pt}\discretionary{.}{%
}{.}\hspace{.4pt}20191187}}


\bibitem{mohseni_trust_2020}
S.~Mohseni, F.~Yang, S.~Pentyala, M.~Du, Y.~Liu, N.~Lupfer, X.~Hu, S.~Ji, and
  E.~D. Ragan.
\newblock Trust {Evolution} {Over} {Time} in {Explainable} {AI} for {Fake}
  {News} {Detection}.
\newblock In {\em Fair \& Responsible AI Workshop at CHI 2020}, 2020.

\bibitem{nourani_investigating_2020}
M.~Nourani, D.~R. Honeycutt, J.~E. Block, C.~Roy, T.~Rahman, E.~D. Ragan, and
  V.~Gogate.
\newblock Investigating the {Importance} of {First} {Impressions} and
  {Explainable} {AI} with {Interactive} {Video} {Analysis}.
\newblock In {\em Extended {Abstracts} of the 2020 {CHI} {Conference} on
  {Human} {Factors} in {Computing} {Systems}}, pp. 1--8. ACM, Honolulu HI USA,
  Apr. 2020. doi: {{%
10\hspace{.1pt}\discretionary{.}{%
}{.}\hspace{.4pt}1145\discretionary{/}{%
}{/}3334480\hspace{.1pt}\discretionary{.}{%
}{.}\hspace{.4pt}3382967}}


\bibitem{nourani_role_2020}
M.~Nourani, J.~King, and E.~Ragan.
\newblock The role of domain expertise in user trust and the impact of first
  impressions with intelligent systems.
\newblock In {\em Proceedings of the AAAI Conference on Human Computation and
  Crowdsourcing}, vol.~8, pp. 112--121, 2020.

\bibitem{papenmeier_how_2019}
A.~Papenmeier, G.~Englebienne, and C.~Seifert.
\newblock How model accuracy and explanation fidelity influence user trust.
\newblock {\em arXiv:1907.12652 [cs]}, July 2019.
\newblock arXiv: 1907.12652.

\bibitem{ribeiro_why_2016}
M.~T. Ribeiro, S.~Singh, and C.~Guestrin.
\newblock "{Why} {Should} {I} {Trust} {You}?": {Explaining} the {Predictions}
  of {Any} {Classifier}.
\newblock In {\em Proceedings of the 22nd {ACM} {SIGKDD} {International}
  {Conference} on {Knowledge} {Discovery} and {Data} {Mining}}, {KDD} '16, pp.
  1135--1144. Association for Computing Machinery, New York, NY, USA, Aug.
  2016. doi: {{%
10\hspace{.1pt}\discretionary{.}{%
}{.}\hspace{.4pt}1145\discretionary{/}{%
}{/}2939672\hspace{.1pt}\discretionary{.}{%
}{.}\hspace{.4pt}2939778}}


\bibitem{sacha_role_2016}
D.~Sacha, H.~Senaratne, B.~C. Kwon, G.~Ellis, and D.~A. Keim.
\newblock The {Role} of {Uncertainty}, {Awareness}, and {Trust} in {Visual}
  {Analytics}.
\newblock {\em IEEE Transactions on Visualization and Computer Graphics},
  22(1):240--249, Jan. 2016. doi: {{%
10\hspace{.1pt}\discretionary{.}{%
}{.}\hspace{.4pt}1109\discretionary{/}{%
}{/}TVCG\hspace{.1pt}\discretionary{.}{%
}{.}\hspace{.4pt}2015\hspace{.1pt}\discretionary{.}{%
}{.}\hspace{.4pt}2467591}}


\bibitem{uggirala_measurement_2004}
A.~Uggirala, A.~K. Gramopadhye, B.~J. Melloy, and J.~E. Toler.
\newblock Measurement of trust in complex and dynamic systems using a
  quantitative approach.
\newblock {\em International Journal of Industrial Ergonomics}, 34(3):175--186,
  Sept. 2004. doi: {{%
10\hspace{.1pt}\discretionary{.}{%
}{.}\hspace{.4pt}1016\discretionary{/}{%
}{/}j\hspace{.1pt}\discretionary{.}{%
}{.}\hspace{.4pt}ergon\hspace{.1pt}\discretionary{.}{%
}{.}\hspace{.4pt}2004\hspace{.1pt}\discretionary{.}{%
}{.}\hspace{.4pt}03\hspace{.1pt}\discretionary{.}{%
}{.}\hspace{.4pt}005}}


\bibitem{yang_how_2020}
F.~Yang, Z.~Huang, J.~Scholtz, and D.~L. Arendt.
\newblock How do visual explanations foster end users' appropriate trust in
  machine learning?
\newblock In {\em Proceedings of the 25th {International} {Conference} on
  {Intelligent} {User} {Interfaces}}, pp. 189--201. ACM, Cagliari Italy, Mar.
  2020. doi: {{%
10\hspace{.1pt}\discretionary{.}{%
}{.}\hspace{.4pt}1145\discretionary{/}{%
}{/}3377325\hspace{.1pt}\discretionary{.}{%
}{.}\hspace{.4pt}3377480}}


\bibitem{yin_understanding_2019}
M.~Yin, J.~Wortman~Vaughan, and H.~Wallach.
\newblock Understanding the {Effect} of {Accuracy} on {Trust} in {Machine}
  {Learning} {Models}.
\newblock In {\em Proceedings of the 2019 {CHI} {Conference} on {Human}
  {Factors} in {Computing} {Systems}}, pp. 1--12. ACM, Glasgow Scotland Uk, May
  2019. doi: {{%
10\hspace{.1pt}\discretionary{.}{%
}{.}\hspace{.4pt}1145\discretionary{/}{%
}{/}3290605\hspace{.1pt}\discretionary{.}{%
}{.}\hspace{.4pt}3300509}}


\bibitem{bernhaupt_effects_2017}
J.~Zhou, S.~Z. Arshad, S.~Luo, and F.~Chen.
\newblock Effects of {Uncertainty} and {Cognitive} {Load} on {User} {Trust} in
  {Predictive} {Decision} {Making}.
\newblock In R.~Bernhaupt, G.~Dalvi, A.~Joshi, D.~K.~Balkrishan, J.~O’Neill,
  and M.~Winckler, eds., {\em Human-{Computer} {Interaction} – {INTERACT}
  2017}, vol. 10516, pp. 23--39. Springer International Publishing, Cham, 2017.
\newblock Series Title: Lecture Notes in Computer Science. doi: {{%
10\hspace{.1pt}\discretionary{.}{%
}{.}\hspace{.4pt}1007\discretionary{/}{%
}{/}978\discretionary{%
}{-}{-}3\discretionary{%
}{-}{-}319\discretionary{%
}{-}{-}68059\discretionary{%
}{-}{-}0\_2}}


\end{thebibliography}
\end{document}